\documentclass[english,1p]{elsarticle}
\usepackage[T1]{fontenc}
\usepackage[latin9]{inputenc}
\usepackage{amsmath}
\usepackage{amssymb}


\usepackage{babel}

\begin{document}

\title{Extending the Spin Projection Operators for Gravity Models with Parity-Breaking
in 3-D}

\author[cbpf]{C. A. Hernaski}

\ead{carlos@cbpf.br}

\author[cbpf]{B. Pereira-Dias}

\ead{bpdias@cbpf.br}

\author[cbpf]{A. A. Vargas-Paredes}

\ead{alfredov@cbpf.br}

\address[cbpf]{Centro Brasileiro de Pesquisas Físicas, Rua Dr. Xavier Sigaud 150,
Urca, Rio de Janeiro, Brazil, CEP 22290-180. Telephone: +55-21-21417254
/ +55-21-21417220}
\begin{abstract}
We propose a new basis of spin-operators, specific for the case of
planar theories, which allows a Lagrangian decomposition into spin-parity
components. The procedure enables us to discuss unitarity and spectral
properties of gravity models with parity-breaking in a systematic
way.\end{abstract}
\maketitle

\section{Introduction}

In the analysis of quantum aspects of any field theory, considerable
interest is devoted to the description of the particle spectrum and
the relativistic and quantum properties of scattering processes of
the theory under investigation. Some of these issues may be understood
by means of the analysis of the propagator of the theory. There are
various methods for the attainment of propagators, but, particularly
in the case of weak field approximation for quantum gravity, which
is our interest, algebraic methods have been intensively developed,
specially the one based on the spin projection operators (SPO). The
SPO has the interesting property of decomposing fields into definite
spin-parity sectors and the latter can be expressed in terms of the
transverse ($\theta$) and longitudinal ($\omega$) operators as building
blocks. The attainment of the propagator by this technique for gravity
models, whenever the metric is adopted as the fundamental quantum
field, was possible using the basis built up in Ref. \citep{Rivers:1964}.
Later Neville \citep{Neville:1978bk}, and Sezgin and Nieuwenhuizen
\citep{Sezgin:1979zf} extended the set of operators in order to provide
a complete SPO basis (in four dimensions) for Lagrangians containing
a rank-2 tensor and a rank-3 tensor antisymmetric in two indices.
With this basis, it was possible to discuss generalized parity-preserving
models of gravitation with the vielbein ($e_{\mu}^{a}$) and spin
connection ($\omega_{\mu}^{\ ab}$) as fundamental fields.

Motivated by the importance of finding a suitable basis in the task
of calculating tensor field propagators, this Letter sets out to propose
and discuss a possible extension of the basis of spin operators mentioned
above \citep{Neville:1978bk,Sezgin:1979zf} that may prove to be more
appropriate for the analysis of propagators of planar models, in special
generalized models for 3-D gravity with parity-breaking.

To understand the convenience of the properties satisfied by the basis
proposed in \citep{Sezgin:1979zf}, let us study a general parity-preserving
model:

\begin{equation}
\left(\mathcal{L}\right)_{2}={\displaystyle \sum_{\alpha,\beta}\psi_{\alpha}O_{\alpha\beta}\psi_{\beta},\label{S-Nlagrangian}}\end{equation}
where $O_{\alpha\beta}$ is a local differential operator and $\psi_{\alpha}$
carry the fundamental quantum fields of the model. We can systematically
analyse the spectrum and unitarity of this model by means of a decomposition
in SPO in the momentum space, as described in \citep{Sezgin:1979zf}:

\begin{equation}
\left(\mathcal{L}\right)_{2}={\displaystyle \sum_{\alpha,\beta,ij,J^{P}}\psi_{\alpha}a_{ij}^{\psi\lambda}\left(J^{P}\right)P_{ij}^{\psi\lambda}\left(J^{P}\right)_{\alpha\beta}\psi_{\beta},}\label{eq:quadratic lagrangian}\end{equation}
where the diagonal operators, $P_{ii}^{\Psi\Psi}\left(J^{P}\right)$,
are projectors in the spin $\left(J\right)$ and parity $\left(P\right)$
sectors of the field $\Psi$ and the off-diagonal operators $(i\neq j)$
implement mappings inside the spin-parity subspace.

This basis of operators is orthonormal and complete in the following
sense:

\begin{subequations} \begin{align}
{\displaystyle \sum_{\beta}P_{ij}^{\Sigma\Pi}\left(J^{P}\right)_{\alpha\beta}P_{kl}^{\Lambda\Xi}\left(I^{Q}\right)_{\beta\gamma}} & =\delta^{PQ}\delta^{\Pi\Xi}\delta_{jk}\delta_{IJ}P_{il}^{\Sigma\Xi}\left(J^{P}\right)_{\alpha\gamma},\label{S-N op basis}\\
{\displaystyle \sum_{i,J^{P}}P_{ii}\left(J^{P}\right)_{\alpha\beta}} & =\delta_{\alpha\beta}.\label{orthonormality relation for the basis}\end{align}
 \end{subequations}

If the coefficient matrices, $a_{ij}(J^{P})$, are invertible, then
the propagator saturated with the external sources, $S_{\alpha}$,
can be written as

\begin{equation}
\Pi=i{\displaystyle \sum S_{\alpha}^{\ast}a_{ij}^{-1\psi\phi}\left(J^{P}\right)P_{ij}^{\psi\phi}\left(J^{P}\right)_{\alpha\beta}S_{\beta}.\label{propagator}}\end{equation}
But, if there are gauge symmetries in the model, the coefficient matrices
become degenerate. In this case, as shown in \citep{Neville:1978bk},
the correct saturated propagator is given by

\begin{equation}
\Pi=i{\displaystyle \sum S_{\alpha}^{\ast}A_{ij}^{\psi\phi}\left(J^{P}\right)P_{ij}^{\psi\phi}\left(J^{P}\right)_{\alpha\beta}S_{\beta},}\label{eq:propagatorwithdegeneracies}\end{equation}
where the $A_{ij}$ are the inverses of the largest sub-matrix with
nonzero determinant obtained from the $a_{ij}$. The sources, in this
case, obey certain constraints. Both, the gauge transformations of
the fields and the source constraints, are obtained from the degeneracy
structures of the coefficient matrices. They are given, respectively,
by:

\begin{subequations} \begin{align}
 & \delta\phi_{\alpha}={\displaystyle \sum_{J^{P},j,\beta,n}V_{j}^{(R,n)}(J^{P})P_{jk}(J^{P})_{\alpha\beta}f_{\beta}(J^{P}),\ \ \text{for any k}}\\
 & {\displaystyle \sum_{j,\beta}V_{j}^{(L,n)}(J^{P})P_{kj}(J^{P})_{\alpha\beta}S_{\beta}(J^{P})=0,\ \ \text{for any }k\text{ and }J^{P}}\end{align}
 \end{subequations} with $f_{\beta}(J^{P})$ being arbitrary functions
and $V^{(R,n)}$ and $V^{(L,n)}$ being the right and left null eigenvectors
of $a_{ij}(J^{P})$. So, they are given by the relations:

\begin{subequations} \begin{align}
 & {\displaystyle \sum_{j}a_{ij}(J^{P})V_{j}^{(R,n)}(J^{P})=0,}\\
 & {\displaystyle \sum_{j}V_{j}^{(L,n)}(J^{P})a_{ji}(J^{P})=0,}\end{align}
 \end{subequations}

We see, by this brief discussion, that with the basis (\ref{S-N op basis}),
(\ref{orthonormality relation for the basis}), the analysis of the
particular model we have at hand can be reduced to the task of discussing
the coefficient matrices. So, it is interesting to generalize this
basis in order to accommodate more general models while keeping the
same type of formalism. Even if this procedure may readily be generalized
to arbitrary dimensions \citep{Hernaski:2009wp}, it may however leave
aside important models with parity violation.

The motivation for our quest comes mainly from the Chern-Simons term
which appears for Yang-Mills and gravity theories in $(1+2)$-dimensional
space-time, that have been extensively discussed in the literature
\citep{Deser:1981wh,HelayelNeto:2010rr,Boldo:1999qw,Nunes:1993yu,Oda:2009im}.
Our point is that the operator brought about by the Chern-Simons term
in a Maxwell-Chern-Simons model (we shall refer to such an operator
as $S_{\mu\nu}$), motivates us to search for operators more fundamental
than the ordinary $\theta_{\mu\nu}$- and $\omega_{\mu\nu}$-operators.
Indeed, we shall find out two new projection operators, $\rho_{\mu\nu}$
and $\sigma_{\mu\nu}$, in terms of which $\theta_{\mu\nu}$ can be
expressed. Our task here consists in building up a whole set of new
SPO in 3-D and, with the help of the results presented in this Letter,
we shall pave the road for the analysis of the spectral consistency
of planar quantum-field theoretic models with vector and tensor fields
that may encompass generalized gravity models in 3-D.

\section{Building up the SPO basis}

To fix ideas before we go on searching for the new basis, it is instructive
to consider a simpler case where the Levi-Civita tensor is present.
In 3-D, we can define the Maxwell-Chern-Simons Lagrangian as:

\begin{equation}
{\cal \mathcal{L}}_{MCS}=-\frac{1}{4}F_{\mu\nu}F^{\mu\nu}+\frac{\mu}{2}\epsilon^{\mu\nu\kappa}A_{\mu}\partial_{\nu}A_{\kappa}.\label{MCS lagrangian}\end{equation}

It is easy to convince ourselves that, if one allows to express the
wave operators only in terms of the metric tensor and derivatives
(powers of momenta in momentum space), the basic elements needed to
expand the operator are the $\theta$'s and $\omega$'s. This is not
the case if the Levi-Civita tensor appears in the wave operator. Since
$\epsilon$ cannot be written in terms of $\theta$'s and $\omega$'s,
we are forced to enlarge the number of building blocks and, this is
actually our main point to extend the usual basis of spin-operators
as we have already mentioned above.

The Lagrangian (\ref{MCS lagrangian}) can be brought into the form:

\begin{equation}
\mathcal{L}_{MCS}=\frac{1}{2}A^{\mu}O_{\mu\nu}A^{\nu},\end{equation}
with $O_{\mu\nu}$ in momentum space, given by:

\begin{equation}
O_{\mu\nu}=\theta_{\mu\nu}+\mu S_{\mu\nu},\end{equation}
where $S_{\mu\kappa}=\epsilon_{\mu\nu\kappa}k^{\nu}$.

If we wish to obtain the propagator, we must know the algebraic properties
between basic operators that we have at hand. We can show that:

\begin{eqnarray}
\theta^{2}=\theta,\ \omega^{2}=\omega,\ \theta\omega=\omega\theta=0,\nonumber \\
S^{2}=-\square\theta,\ S\theta=\theta S=S,\ S\omega=\omega S=0.\end{eqnarray}

With these relations, we see that the operator $S$ is a transverse
one. That is, it is a linear operator that maps an arbitrary vector
into another vector that lies in the transverse subspace. But, this
vector is independent of the vector that is obtained by the action
of the $\theta$-operator. This is possible since the transverse subspace
in 3-D is two dimensional. Surely, we can exhaust all possible transverse
operators if we define a basis in this transverse subspace. Taking
two orthonormal space-like vectors ($e_{1}$ and $e_{2}$) in the
transverse subspace, we may define two operators that project onto
the one-dimensional subspace spanned by each one of these vectors
and two operators that implement mappings between these two subspaces.
Let us define the two projectors by the relation:

\begin{equation}
\theta_{\mu\nu}=\rho_{\mu\nu}+\sigma_{\mu\nu},\label{teta in terms of ro and sigma}\end{equation}
 with

\begin{subequations} \begin{align}
\rho_{\mu\nu} & =-(e_{1})_{\mu}(e_{1})_{\nu},\label{eq:rho}\\
\sigma_{\mu\nu} & =-(e_{2})_{\mu}(e_{2})_{\nu},\label{eq:sigma}\end{align}
 \end{subequations} where,

\begin{equation}
e_{1}.e_{1}=e_{2}.e_{2}=-1,e_{1}.e_{2}=0.\end{equation}

One can show that the other two operators that accomplish the mappings
can be given by:

\begin{subequations} \begin{align}
(P_{12})_{\mu\nu} & =\epsilon^{\rho\sigma\lambda}\rho_{\mu\rho}\sigma_{\nu\sigma}\bar{k}_{\lambda},\label{eq:P12}\\
(P_{21})_{\mu\nu} & =\epsilon^{\rho\sigma\lambda}\sigma_{\mu\sigma}\rho_{\nu\rho}\bar{k}_{\lambda}.\label{eq:P21}\end{align}
 \end{subequations}where $\bar{k}_{\lambda}=\frac{k_{\lambda}}{\sqrt{k^{2}}}$.
These four operators satisfy orthogonality conditions: $(P_{11})^{2}=P_{11}$,
$(P_{22})^{2}=P_{22},$ $P_{12}P_{21}=P_{11},$ and $P_{21}P_{12}=P_{22},$
with $P_{11}\equiv\rho$ and $P_{22}\equiv\sigma$.

The task of finding a basis of operators that act on the vectors fields
$\Lambda$ has already been carried out, since we only need to add
the longitudinal operator, $\omega$, to the operators \eqref{eq:rho},
\eqref{eq:sigma}, \eqref{eq:P12}, and \eqref{eq:P21}. In the work
of Ref.\citep{Hernaski:2009wp}, the spin projectors for symmetric
rank-2 tensor was obtained for arbitrary dimension. These projectors
have been written in terms of $\theta$'s and $\omega$'s. But, as
we have seen, $\theta$ can be split into two more basic projectors
and, with this, we increase the possibilities of construction of wave
operators. In the same way, we can also use the relation (\ref{teta in terms of ro and sigma})
to split the spin projectors of \citep{Hernaski:2009wp} for D=3 into
more basic ones. As an example, let us take one of the projectors
and analyse how this works:

\begin{equation}
P^{\Psi\Psi}(2^{+})_{ab;cd}=\frac{1}{2}(\theta_{ac}\theta_{bd}+\theta_{ad}\theta_{bc})-\frac{1}{2}\theta_{ab}\theta_{cd}.\label{spin projector}\end{equation}
Substituting (\ref{teta in terms of ro and sigma}) in the expression
(\ref{spin projector}), we obtain two projectors in terms of $\rho$
and $\sigma$, one for each degree of freedom of spin:

\begin{subequations} \begin{align}
P_{11}^{\Psi\Psi}(2^{-})_{ab;cd} & =\frac{1}{2}(\rho_{ac}\sigma_{bd}+\rho_{ad}\sigma_{bc}+\sigma_{ac}\rho_{bd}+\sigma_{ad}\rho_{bc}),\label{ro-sigma-spin-2-11}\\
P_{22}^{\Psi\Psi}(2^{+})_{ab;cd} & =\frac{1}{2}(\rho_{ad}\rho_{bc}+\sigma_{ad}\sigma_{bc})-\frac{1}{2}(\rho_{ab}\sigma_{cd}+\sigma_{ab}\rho_{cd}).\label{ro-sigma-spin-2-22}\end{align}
 \end{subequations} The mappings between the degrees of freedom are
carried out by:

\begin{subequations} \begin{align}
P_{12}^{\Psi\Psi}(2^{-+})_{ab;cd} & =\frac{1}{2}\epsilon_{ghe}(\rho_{ac}\sigma_{b}^{h}\rho_{d}^{g}+\rho_{bc}\sigma_{a}^{h}\rho_{d}^{g}-\sigma_{ad}\rho_{b}^{g}\sigma_{c}^{h}-\sigma_{bd}\rho_{a}^{g}\sigma_{c}^{h})\bar{k}^{e},\label{ro-sigma-spin-2-12}\\
P_{21}^{\Psi\Psi}(2^{+-})_{ab;cd} & =\frac{1}{2}\epsilon_{ghe}(\rho_{ca}\sigma_{d}^{h}\rho_{b}^{g}+\rho_{da}\sigma_{c}^{h}\rho_{b}^{g}-\sigma_{cb}\rho_{d}^{g}\sigma_{a}^{h}-\sigma_{bd}\rho_{c}^{g}\sigma_{a}^{h})\bar{k}^{e}.\label{ro-sigma-spin-2-21}\end{align}
 \end{subequations} Before we proceed, let us clarify the notation.
The notation in (\ref{spin projector}) is imported from 4-D and it
makes strictly physical sense only in 4-D. If the symbols do not lead
to wrong physical conclusions, we preserve them in 3-D. But, in terms
of $\rho$ and $\sigma$, extra care must be taken. Actually, the
operators above do not project over the whole spin-2 space, but rather
over a sub-sector of the degrees of freedom carried by a spin-2. The
most important difference concerns parity. In 4-D, we can fix the
parity of an operator by counting the number of field contractions
with the $\theta$'s present in the given operator. This is so because
$\theta$ projects a Lorentz index in the $1^{-}$-sector. That is,
we associate a parity \textquotedbl{}-\textquotedbl{} with the subspace
projected by $\theta$. This makes sense, since the representation
of parity in Minkowski vector space is given by:

\begin{equation}
P=\left(\begin{array}{cccc}
1 & 0 & 0 & 0\\
0 & -1 & 0 & 0\\
0 & 0 & -1 & 0\\
0 & 0 & 0 & -1\end{array}\right),\end{equation}
 and, for a massive particle in the rest frame, we can assume that
the transverse space is the 3-D spatial part of Minkowski space. So,
the parity operation changes the sign of the spatial components of
the vector. However, in 3-D, a parity operator distinguishes one particular
space direction. For example, we can define it as: \begin{equation}
P=\left(\begin{array}{ccc}
1 & 0 & 0\\
0 & 1 & 0\\
0 & 0 & -1\end{array}\right).\end{equation}
 In this form, the transverse operator can be split as the direct
sum of two subspaces, each one associated with one parity. By convention,
let us choose the subspace projected by $\sigma$ as the one related
to the \textquotedbl{}-\textquotedbl{} parity. So, in 3-D, the parity
of the operators is given by counting the number of indices contracted
by the $\sigma$ operator. This justifies the prescription we have
done to the operators (\ref{ro-sigma-spin-2-11}), (\ref{ro-sigma-spin-2-22}),
(\ref{ro-sigma-spin-2-12}) and (\ref{ro-sigma-spin-2-21}).

By construction, the operators defined above satisfy: \begin{align}
P^{\Psi\Psi}(2^{+}) & =P_{11}^{\Psi\Psi}(2^{++})+P_{22}^{\Psi\Psi}(2^{--}),\end{align}
 and orthogonality relation: $(P_{11})^{2}=P_{11}$, $(P_{22})^{2}=P_{22},$
$P_{12}P_{21}=P_{11},$ and $P_{21}P_{12}=P_{22}$.

This process of decomposition can be repeated for all operators needed
to exhaust all the possibilities of contraction of the fields in the
free Lagrangian. Before we write down the explicit form of the operators
in our basis, it is worthy to mention that they carry a pair of superscripts
$\Psi$ and $\Lambda$. $\Psi$ denotes a set of rank two tensors
and $\Lambda$ denotes set of vector fields, which depends on the
set of fields in each specific model. The parity of each operators
can be read by the sign cast in the matrices. We finally cast the
operators, in terms of $\rho$, $\sigma$ and $\omega$, below:\begin{align}
P(0) & =\negthickspace\negthickspace\begin{array}{c}
\Psi\\
\Psi\\
\Lambda\end{array}\negthickspace\negthickspace\overset{\Psi\qquad\ \ \ \ \ \ \ \ \ \Psi\qquad\ \ \ \ \ \ \ \ \ \Lambda}{\left[\begin{array}{ccc}
\frac{1}{2}\theta_{ab}\theta_{cd} & \frac{1}{\sqrt{2}}\theta_{ab}\omega_{cd} & \frac{1}{\sqrt{2}}\theta_{ab}\bar{k}_{c}\\
\frac{1}{\sqrt{2}}\omega_{ab}\theta_{cd} & \omega_{ab}\omega_{cd} & \omega_{ab}\bar{k}_{c}\\
\frac{1}{\sqrt{2}}\theta_{bc}\bar{k}_{a} & \omega_{bc}\bar{k}_{a} & \omega_{ab}\end{array}\right]}\end{align}
 \begin{equation}
P(1)=\negthickspace\negthickspace\begin{array}{c}
\Psi\left(+\right)\\
\Psi\left(-\right)\\
\Lambda\left(+\right)\\
\Lambda\left(-\right)\end{array}\negthickspace\negthickspace\overset{\Psi\left(+\right)\qquad\qquad\qquad\Psi\left(-\right)\qquad\qquad\qquad\Lambda\left(+\right)\qquad\qquad\qquad\Lambda\left(-\right)}{\left[\negthickspace\negthickspace\begin{array}{cccc}
2\rho_{ac}\omega_{bd}\negthickspace & \negthickspace2\epsilon_{ghe}\rho_{a}^{g}\sigma_{c}^{h}\omega_{bd}\bar{k}^{e}\negthickspace & \negthickspace\sqrt{2}\rho_{ac}\bar{k}_{b}\negthickspace & \negthickspace\sqrt{2}\epsilon_{ghe}\rho_{ag}\sigma_{c}^{h}\omega_{be}\\
2\epsilon_{ghe}\sigma_{a}^{h}\rho_{c}^{g}\omega_{db}\bar{k}^{e}\negthickspace & \negthickspace2\sigma_{ac}\omega_{bd}\negthickspace & \negthickspace\sqrt{2}\epsilon_{ghe}\sigma_{a}^{h}\rho_{c}^{g}\omega_{eb}\negthickspace & \negthickspace\sqrt{2}\sigma_{ac}\bar{k}_{b}\\
\sqrt{2}\rho_{ba}\bar{k}_{c}\negthickspace & \negthickspace\sqrt{2}\epsilon_{ghe}\sigma_{b}^{h}\rho_{a}^{g}\omega_{ec}\negthickspace & \negthickspace\rho_{ab}\negthickspace & \negthickspace\epsilon^{fgh}\rho_{af}\sigma_{bg}\bar{k}_{h}\\
\sqrt{2}\epsilon_{ghe}\rho_{bg}\sigma_{a}^{h}\omega_{ce}\negthickspace & \negthickspace\sqrt{2}\sigma_{ba}\bar{k}_{c}\negthickspace & \negthickspace-\epsilon^{def}\sigma_{ad}\rho_{be}\bar{k}_{f}\negthickspace & \negthickspace\sigma_{ab}\end{array}\negthickspace\negthickspace\right]}\end{equation}
\begin{equation}
P(2)=\negthickspace\negthickspace\begin{array}{c}
\Psi\left(+\right)\\
\Psi\left(-\right)\end{array}\negthickspace\negthickspace\overset{\qquad\Psi\left(+\right)\qquad\qquad\qquad\qquad\qquad\qquad\qquad\Psi\left(-\right)}{\left[\negthickspace\negthickspace\begin{array}{cc}
2\rho_{ac}\sigma_{bd} & \epsilon_{ghe}(\rho_{ca}\sigma_{d}^{h}\rho_{b}^{g}-\sigma_{cb}\rho_{d}^{g}\sigma_{a}^{h})\bar{k}^{e}\\
\epsilon_{ghe}(\rho_{ac}\sigma_{b}^{h}\rho_{d}^{g}-\sigma_{ad}\rho_{b}^{g}\sigma_{c}^{h})\bar{k}^{e} & \frac{1}{2}(\rho_{ad}\rho_{bc}+\sigma_{ad}\sigma_{bc}-\rho_{ab}\sigma_{cd}-\sigma_{ab}\rho_{cd})\end{array}\negthickspace\negthickspace\right]}\end{equation}

Also, it is understood the operators share the same symmetrization
properties (with numerical factor) of the associated fields.

The off-diagonal operators have been obtained in such a way that the
following multiplicative rules and completeness relation are fulfilled:
\begin{subequations} \begin{align}
{\displaystyle \sum_{\beta}P_{ij}^{\Sigma\Pi}(I^{PQ})_{\alpha\beta}P_{kl}^{\Omega\Xi}(J^{RS})_{\beta\gamma}} & =\delta_{jk}\delta^{\Pi\Omega}\delta^{IJ}\delta^{QR}P_{il}^{\Sigma\Xi}(I^{PS})_{\alpha\gamma},\\
{\displaystyle \sum_{i,I^{PP}}P_{ii}(I^{PP})_{\alpha\beta}} & =\delta_{\alpha\beta},\end{align}
 \end{subequations} and, as we have claimed at the beginning, this
makes possible to analyse generalized parity-violating gravity models
in 3-D, by using the same techniques as the ones presented in \citep{Sezgin:1979zf}.
There are only slight differences due to the notation and role played
by parity. In the present case, the wave operators is written as:

\begin{equation}
O_{\alpha\beta}={\displaystyle \sum_{J,ij}a_{ij}^{\Sigma\Pi}(J)P_{ij}^{\Sigma\Pi}(J^{PQ})_{\alpha\beta},}\label{eq:waveOperatorParityBreaking}\end{equation}
 and the saturated propagator, in the case of gauge symmetries, can
be cast as below:

\begin{equation}
\Pi=i{\displaystyle \sum_{J,ij}S_{\alpha}^{*}A_{ij}^{\Sigma\Pi}(J)P_{ij}^{\Sigma\Pi}(J^{PQ})_{\alpha\beta}S_{\beta},}\label{eq:propagatorDegenaraciesParityBreaking}\end{equation}
where $A_{ij}(J)$ is the inverse of the largest sub-matrix of the
$a_{ij}(J)$ with the degeneracies extracted. The important fact is
that these coefficient matrices accommodate the coefficients of the
operators with both parities. Besides these subtle aspects, the rest
of the analysis goes along the same paths as it has been carried out
with the basis (\ref{S-N op basis}).

\section{Application}

In order to explicitly illustrate how to apply the proposed basis,
we discuss the unitarity properties of a gravity model in the second-order
formalism including higher derivatives and the parity-breaking Chern-Simons
term. The Lagrangian we consider reads as below:

\begin{equation}
\mathcal{L}=\sqrt{g}\left(\alpha R+\beta R_{\mu\nu}R^{\mu\nu}+\gamma R^{2}\right)+\frac{\mu}{2}\mathcal{L}_{CS},\label{eq:LagrangianApplication}\end{equation}
where\begin{equation}
\mathcal{L}_{CS}=\varepsilon^{\mu\nu\kappa}\Gamma_{\mu\rho}^{\sigma}\left(\partial_{\nu}\Gamma_{\kappa\sigma}^{\rho}+\Gamma_{\nu\sigma}^{\lambda}\Gamma_{\kappa\lambda}^{\rho}\right),\end{equation}
and $\alpha$, $\beta$, $\gamma$, and $\mu$ are arbitrary parameters.

After adopting the well-known weak field approximation for the metric:
$g_{\mu\nu}=\eta_{\mu\nu}+h_{\mu\nu}$, the wave operator of the Lagrangian
\eqref{eq:LagrangianApplication} can be brought into the form \eqref{eq:waveOperatorParityBreaking}.
Due to the gauge symmetries of the model, the spin-$0$ matrix is
non-invertible. Then, the propagator is given by \eqref{eq:propagatorDegenaraciesParityBreaking},
where the spin matrices $A_{ij}^{\Sigma\Pi}(J)$ are cast as:

\begin{equation}
A(0)=\frac{2}{p^{2}\left(\left(3\beta+8\gamma\right)p^{2}-\alpha\right)},\label{eq:spin 0}\end{equation}
\begin{equation}
A\left(2\right)=\frac{4}{p^{2}\left[\left(\alpha+\beta p^{2}\right)^{2}-\mu^{2}\right]}\left(\begin{array}{cc}
\frac{1}{2}\left(\alpha+\beta p^{2}\right) & \frac{i}{2}\mu\sqrt{p^{2}}\\
-\frac{i}{2}\mu\sqrt{p^{2}} & \frac{1}{2}\left(\alpha+\beta p^{2}\right)\end{array}\right).\label{eq:spin2}\end{equation}
The conditions for absence of ghosts and tachyons are respectively
given by:

\begin{equation}
\Im\text{Res}(\Pi|_{p^{2}=m^{2}})>0,\quad\mbox{and}\quad m^{2}\geq0.\label{eq:ghost/tachyon conditions for massive poles}\end{equation}

The condition for absence of ghosts for each spin is directly related
to the positivity of the matrices $\left(\sum A_{ij}\left(J,m^{2}\right)P_{ij}\right)_{\alpha\beta}$,
where $A_{ij}\left(J,m^{2}\right)\equiv\mbox{Res}\left.\left(A_{ij}\left(J\right)\right)\right|_{p^{2}=m^{2}}$.
However, it can be shown that these matrices have only one non-vanishing
eigenvalue at the pole, which is equal to the trace of $A(J,m^{2})$.
Also, the operators $P_{ij}$ themselves contribute only with a sign
$(-1)^{N}$, whenever calculated at the pole, where $N$ is the sum
of the number of $\rho$'s and $\sigma$'s in each term of the projector.
Therefore, the condition for absence of ghosts among the massive modes
for each spin takes a simple final form:

\begin{equation}
(-1)^{N}\mbox{tr}A(J,m^{2})|_{p^{2}=m^{2}}>0.\label{eq:ghost condition in terms of the trace}\end{equation}

Using the constraints \eqref{eq:ghost/tachyon conditions for massive poles}
and \eqref{eq:ghost condition in terms of the trace} for the matrices
\eqref{eq:spin 0}-\eqref{eq:spin2}, so that ghosts and tachyons
be absent, we get the following conditions for the parameters:

\begin{subequations}

\begin{eqnarray}
 &  & \mbox{Spin-}\mathbf{2}:\ \alpha<0,\ \ \beta>0;\\
 &  & \mbox{Spin-}\mathbf{0}:\ \alpha>0,\ \ 3\beta+8\gamma>0.\end{eqnarray}

\end{subequations}

For arbitrary values of the parameters, the model is therefore non-unitary.
One way to circumvent this problem is to inhibit the propagation of
the massive spin- $0$ mode, by taking $3\beta+8\gamma=0$. Remarkably,
this is exactly the condition considered in the Bergshoeff-Hohm-Townsend
(BHT) model\citep{Bergshoeff:2009hq}. 

For the massless poles, extra care must be taken. Since the parity-operators
are singular at the massless pole, one must use the original expression
\eqref{eq:propagatorDegenaraciesParityBreaking} for the propagator
in order to compute the residue. The constraints satisfied by the
sources allow us to handle correctly the singularities. Using such
constraints and discarding terms that do not contribute to the residue,
one obtains:

\begin{equation}
\Pi=\frac{2}{\alpha p^{2}}i\tau^{\ast ab}\left(\left(\frac{1}{2}\left(\eta_{ac}\eta_{bd}+\eta_{ad}\eta_{bc}\right)-\eta_{ab}\eta_{cd}\right)+i\mu\varepsilon^{aec}\eta^{bd}p_{e}\right)\tau^{cd}.\end{equation}

With a suitable basis for the sources in momentum space, one can show
that this expression vanishes identically and for this reason there
is no propagating massless mode. With these results, we conclude,
as well-known, that the BHT-model is unitary. We hope this discussion
has been useful to illustrate the use of our basis of operators.

\section{Concluding Remarks}

In this paper, we have proposed a orthonormal basis of operators suitable
to carry out the task of deriving the propagators of models that can
include parity-breaking terms. The presence of the Levi-Civita symbol
in these terms suggests a convenient splitting of the degrees of freedom
of the fields in terms of parity-components rather than spin-components.
Since every massive particle in three dimensions has two helicities
in spite of its spin, the decomposition in parity-components yields
in a splitting in individual degrees of freedom. This basis is only
defined for time-like momenta, such as the usual operators in terms
of $\theta's$ and $\omega's$ . However, the singularities that appears
for light-like momenta can be consistently handled with the fully
saturated propagators, as we have explicitly done in our application
example. The consistency of the results obtained with this basis has
been tested in a well-known model, viz., the BHT-model. We reproduce
in this paper the conditions for the unitarity for this model.

The systematic way of analysing the spectrum consistency can now be
readily implemented for other parity-breaking-type models. Interesting
ones are those related to the Lorentz-breaking gravity models in four
dimensions. For example, action terms like $\epsilon^{\mu\kappa\lambda}T_{\kappa\lambda}^{\ \ a}R_{\mu a}$,
$R\epsilon^{\mu\nu\kappa}T_{\mu\nu\kappa}$ , $\epsilon^{\mu\nu\kappa}T_{\kappa a}^{\ \ a}R_{\mu\nu}$
could be considered in 3-D as descents from the Lorentz-symmetry breaking
terms for a special choice of the background. Another clear application
of such a basis could appear in connection with parity-conserving
models but taking advantage from the dual aspect of the fields. Fist-order
formulation of gravity in 3-D is a good example where this could happen,
since one can write the quantum fluctuations of the vielbein $e_{\mu}^{a}$
and spin-connection $\omega_{\mu}^{\ ab}$ as follows:

\begin{subequations} \begin{align}
\tilde{e}_{\mu\nu} & =\phi_{\mu\nu}+\epsilon_{\mu\nu\kappa}\chi^{\kappa},\label{duality for e}\\
\tilde{\omega}_{\mu}^{\ \nu\kappa} & =\epsilon^{\nu\kappa\sigma}\left(\psi_{\mu\sigma}+\epsilon_{\mu\sigma\rho}\lambda^{\rho}\right),\label{duality for w}\end{align}
 \end{subequations} $\phi_{\mu\nu}$ is the symmetric part of the
vielbein fluctuation and $\chi^{\kappa}$ is the vector dual to the
antisymmetric one, $\psi_{\mu\sigma}$ is the symmetric part of the
field dual to the spin connection fluctuation and $\lambda^{\rho}$
is the vector dual to the antisymmetric part of the dual field. Indeed,
this study of a 3-D model for gravity in the presence of dynamical
torsion and higher powers of the curvature along with a Chern-Simons
term is under progress and the efficacy of the projectors we have
presented here becomes manifest in this application. These results
shall soon be reported elsewhere.

\textbf{ACKNOWLEDGMENTS:}

The authors express their gratitude to Prof. J. A. Helayël-Neto for
the supporting discussions and for the encouragement for pursuing
this investigation. Prof. S. A. Dias is also acknowledged for helpful
comments and suggestions. Thanks are also due to CNPq-Brazil and FAPERJ
for our Graduate fellowships.

\bibliographystyle{elsarticle-num}
\bibliography{referencias}

\begin{thebibliography}{10}
\expandafter\ifx\csname url\endcsname\relax
  \def\url#1{\texttt{#1}}\fi
\expandafter\ifx\csname urlprefix\endcsname\relax\def\urlprefix{URL }\fi
\expandafter\ifx\csname href\endcsname\relax
  \def\href#1#2{#2} \def\path#1{#1}\fi

\bibitem{Rivers:1964}
R.~J. Rivers, Il Nuovo Cimento 34 (1964) 387.

\bibitem{Neville:1978bk}
D.~E. Neville, Phys. Rev. D18 (1978) 3535.

\bibitem{Sezgin:1979zf}
E.~Sezgin, P.~van Nieuwenhuizen, Phys. Rev. D21 (1980) 3269.

\bibitem{Hernaski:2009wp}
C.~A. Hernaski, A.~A. Vargas-Paredes, J.~A. Helayel-Neto, Phys. Rev. D80 (2009)
  124012.
\newblock \href {http://arxiv.org/abs/0905.1068} {\path{arXiv:0905.1068}}.

\bibitem{Deser:1981wh}
S.~Deser, R.~Jackiw, S.~Templeton, Ann. Phys. 140 (1982) 372--411.

\bibitem{HelayelNeto:2010rr}
J.~A. Helayel-Neto, L.~M. Moraes, V.~J. Vasquez\href
  {http://arxiv.org/abs/1002.0526} {\path{arXiv:1002.0526}}.

\bibitem{Boldo:1999qw}
J.~L. Boldo, L.~M. de~Moraes, J.~A. Helayel-Neto, Class. Quant. Grav. 17 (2000)
  813--823.
\newblock \href {http://arxiv.org/abs/hep-th/9903127}
  {\path{arXiv:hep-th/9903127}}.

\bibitem{Nunes:1993yu}
F.~C.~P. Nunes, G.~O. Pires, Phys. Lett. B301 (1993) 339--344.

\bibitem{Oda:2009im}
I.~Oda, {Renormalizability of Topologically Massive Gravity, }\href
  {http://arxiv.org/abs/0905.1536} {\path{arXiv:0905.1536}}.

\bibitem{Bergshoeff:2009hq}
E.~A. Bergshoeff, O.~Hohm, P.~K. Townsend, Phys. Rev. Lett. 102 (2009) 201301.
\newblock \href {http://arxiv.org/abs/0901.1766} {\path{arXiv:0901.1766}}.

\end{thebibliography}

\end{document}